# An estimate of the circulation generated by a bluff body

T. S. Morton

Department of Mechanical, Aerospace & Biomedical Engineering, University of Tennessee Space Institute
411 B.H. Goethert Parkway, Tullahoma, TN 37388, USA

A loss in circulation is sometimes cited in connection with bluff-body wakes as a result of comparing the circulation actually observed downstream with a well-known theoretical estimate of the total circulation generated by a cylinder. In an effort to better understand this reported loss in circulation, an alternative estimate of the circulation generated by a cylinder is derived by integrating the velocity on a closed loop containing the attached boundary layer. Predictions of the dimensionless circulation for a cylinder in cross flow are less than the previous theoretical estimate and agree with observed values. This suggests that the total circulation generated by bluff bodies may have been overestimated in the past, and that comparison of observed values with this overestimate is the origin of the perceived "loss" in circulation.

## 1. Introduction

The purpose of this study is to provide an estimate of the mean circulation generated by a bluff body in steady motion through a fluid. This circulation has been estimated by considering it to be generated at a rate of $(1 - C_{pb})v_\infty^2/2$ per shedding period (Roshko 1954). Here, $C_{pb}$ is the base pressure coefficient, and $v_\infty$ is the free stream velocity. The following dimensionless form of this estimate was later used in a study of square section cylinders (Bearman & Obasaju 1982):

$$\frac{\Gamma}{Dv_\infty} = \frac{(1 - C_{pb})}{2\,St}, \qquad (1)$$

where $\Gamma$ is the total circulation generated by the body, and $D$ is the body dimension. This estimate predicts a total circulation of more than twice that typically found in bluff-body wakes, and it is generally believed that half or more of the vorticity is cancelled in the near wake due to mixing (see Bearman & Obasaju 1982). The fact that Fage and Johansen (1927) were only able to account for roughly 60% of the total vorticity nine body lengths downstream, which led them to first suggest the mixing of positive and negative vorticity in the wake, appears to be the origin of the justification of (1). They later (Fage & Johansen 1928) made detailed graphical estimates in the near wake, however, and appear to have adequately accounted for the total vorticity to within 10% error. Subsequent studies have shown that only 35~45% of the total vorticity predicted by (1) can be found in the near wake (Cantwell & Coles 1983; Lyn *et al.* 1995). Davies (1976) found that only 26% of the vorticity predicted by (1) is found in the wake of a D-shaped cylinder 8 diameters downstream. Wu *et al.* (1994) reported a value of $\Gamma_{KM}/(Dv_\infty) = 3.5$ for the mean circulation of Kármán vortices behind a cylinder at $Re = 525$. Likewise, in the range $73 \leq Re \leq 226$, a comparable value of $\Gamma_{KM}/(\pi Dv_\infty) \approx 1$ is observed in the cylinder wake (Green & Gerrard 1993). For $Re = 1.44 \times 10^5$, Cantwell & Coles (1983) measured dimensionless circulation values ranging from 2.07 to 2.55 in the wake of a cylinder, whereas (1) gave an estimate of $\Gamma/(Dv_\infty) = 5.86$. Tanaka and Murata (1986) measured a value of $\Gamma/(Dv_\infty) = 2.1$ in the wake of a cylinder at $Re = 3.7 \times 10^4$. The total dimensionless circulation of a rectangular cylinder with a blunt leading edge was likewise found to be approximately 2.0 (Wood 1967). In a computational study of the unsteady wake of a disk, Johari and Stein (2002) obtained a value of $\Gamma/(Dv_\infty) = 2.6$, which compares well with the experimental value of $\Gamma/(Dv_\infty) = 2.5$ (Balligand 2000).

The anomalous mixing of positive and negative vorticity has been characterized as "a long standing puzzle" (Zdravkovich 1989). In what follows, a simple estimate of the dimensionless circulation is derived using a contour integration that includes the attached boundary layer on a circular cylinder in crossflow. Predictions are significantly lower than those of (1) and agree fairly well with values cited above. Consequently, the "anomalous mixing" argument is not required in order to reconcile the present estimation method with the circulation observed in wakes.

## 2. Cylinder Wake Circulation

Due to the no slip condition of the attached boundary layer, a bluff body such as a circular cylinder transmits a torque to the fluid passing it. This torque sets some region of the fluid in motion rotationally. The resulting circulation can be determined by performing the following contour integration

$$\Gamma_{(1/2)} = \oint \boldsymbol{v} \cdot d\boldsymbol{s} \qquad (2)$$

along the loop $\gamma_1 + \gamma_2 + \gamma_3 + \gamma_4$ shown in Figure 1. The subscript "(1/2)" in (2) indicates that this represents only half of the total circulation generated by a symmetric mean flow about the cylinder. The contour integration is performed in 4 parts. On the curve $\gamma_1$ (see Figure 1), the approaching fluid decelerates and stagnates according to the velocity distribution for ideal fluid flow past a cylinder ($n = 2$) or a sphere ($n = 3$). Along $\gamma_2$, which terminates at the separation point, the velocity is zero due to the no-slip condition. The time-mean velocity on the contour $\gamma_3$, which emanates from the separation point, can be found from wake measurements, and its asymptotic dimensionless radial component will be $\cos\theta$, where $\theta$ is the separation angle. The distance from the center of the cylinder to the contour $\gamma_4$ in Figure 1 will be denoted by $R_\infty$. On this contour, the magnitude of the velocity is $v_\infty$, so its clockwise component along the contour $\gamma_4$ is $v_\infty \sin\varphi$, where $\varphi$ is the angular coordinate (measured from the same direction from which $\theta$ is measured) of a given fluid particle on the contour $\gamma_4$.

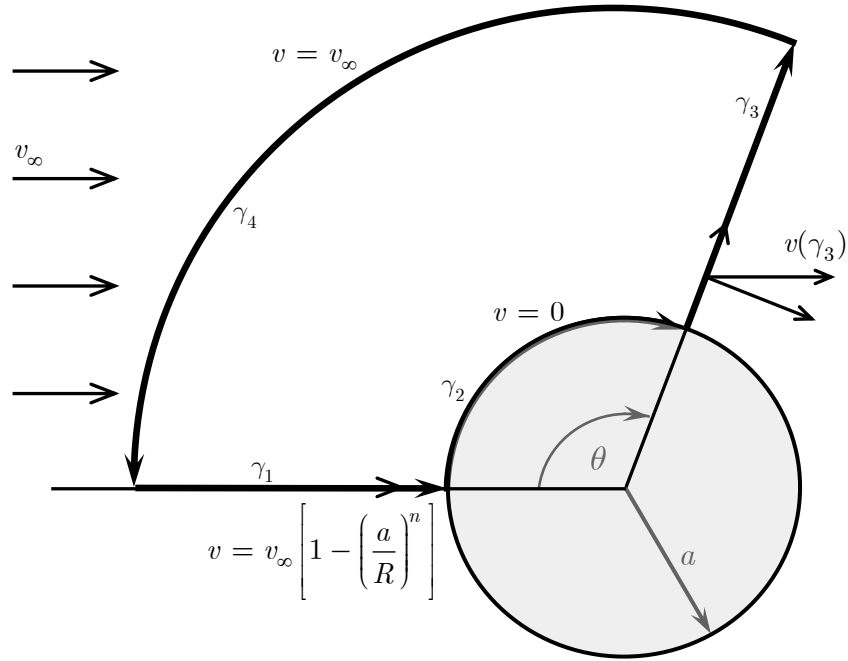

Figure 1. Schematic of integration contour for computing circulation.

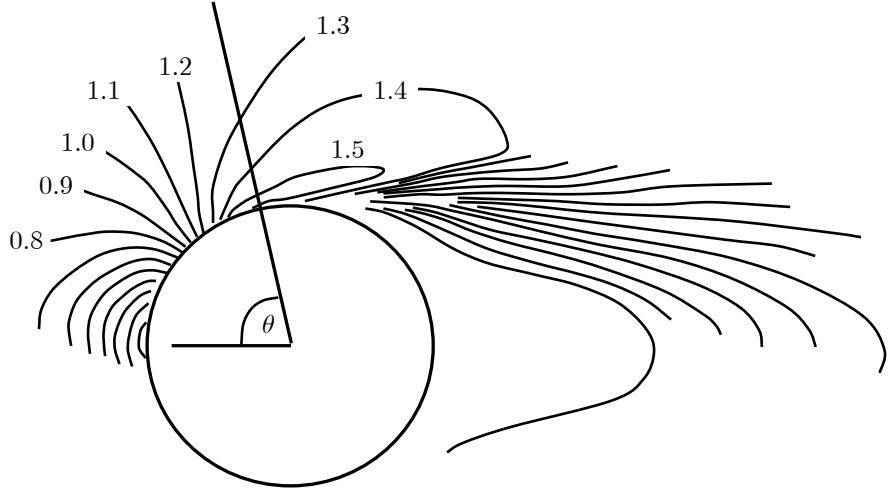

Figure 2. Iso contours of mean streamwise velocity, $v_x$, with separation angle indicated (Adapted from Perrin *et al.* 2006. Cantwell and Coles 1983 reported $\theta = 77°$).

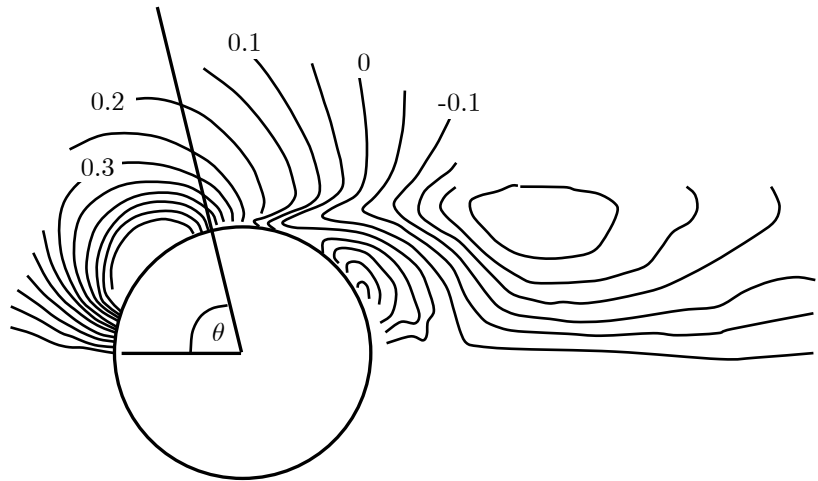

Figure 3. Iso contours of mean vertical velocity, $v_y$, with separation angle indicated. (Adapted from Perrin *et al.* 2006).

The case to be examined herein is the recent study by Perrin *et al.* (2006) of flow past a circular cylinder at $Re = 1.4 \times 10^5$. This is essentially the same Reynolds number tested by Cantwell and Coles (1983); however, the configuration tested by Perrin *et al.* was one in which the aspect ratio of the cylinder was lower and the blockage higher. The separation angle was not reported by Perrin *et al.*, but it was likely comparable to that reported by Cantwell and Coles, namely 77°. The measurements made by Perrin *et al.* also show the mean wake bubble in much clearer detail. The pertinent measurements from that study are sketched in Figures 2 and 3. Values for the velocity components $v_x$ and $v_y$ can be found by locating the intersections of the contours with the separation line (shown drawn at $\theta = 77°$ in the figures). From these velocity components, the profile of the velocity magnitude along the separation line was reconstructed. The resulting mean velocity vectors are shown in Figure 4 emanating from the separation line. (Arrows are not drawn at the terminating ends of the vectors.)

Integration of (2) along contour $\gamma_3$ requires the projection of the velocity in the direction of the contour. This projection is shown in Figure 4 at the base of each velocity vector. Obviously, these projections are very small near the body but grow larger farther from the body. Far from the body, as the magnitude of the dimensionless velocity reaches its asymptotic value of unity and the direction of the vectors shown in Figure 4 becomes horizontal, their components projected in the direction of the contour $\gamma_3$ approach $\cos\theta$. These projected components are plotted in Figure 5. The curve shown extrapolating to the asymptotic value is:

$$\frac{\boldsymbol{v}(\gamma_3)}{v_\infty} \cdot \frac{\mathrm{d}\boldsymbol{s}}{\mathrm{d}s} = \cos\theta - ce^{-m(R-a)/a} + (c - \cos\theta)e^{-k(R-a)/a}. \tag{3}$$

Here, $a$ is the radius of the cylinder, and the constants $c$, $k$, and $m$ are positive. This velocity profile will be used in the integral in (2) to compute the circulation.

Referring to Figure 1, the integration in (2) can now be rewritten piecewise, as follows:

$$\frac{\Gamma_{(1/2)}}{v_\infty} = -\int_{R_\infty}^{a}\left[1 - \left(\frac{a}{r}\right)^n\right]\mathrm{d}r + \int_{a}^{R_\infty}\left(\cos\theta - ce^{-m(r-a)/a} + (c - \cos\theta)e^{-k(r-a)/a}\right)\mathrm{d}r \\ + \int_{\theta}^{0} R_\infty \sin\varphi\, \mathrm{d}\varphi \tag{4}$$

The first term on the right side is negative because outward radial velocity is positive. The last term on the right side is positive because the angular velocity is clockwise and coincides with the direction of increasing $\theta$.

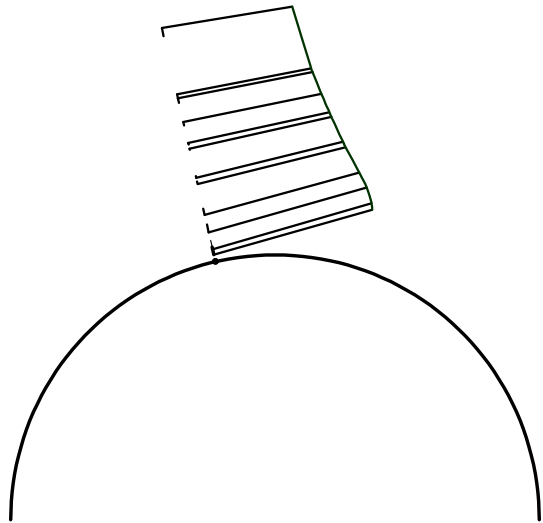

Figure 4. Fluid velocity vectors (drawn without arrows) along the line showing the separation angle. At the base of these vectors, velocity projections along the integration contour are also shown. (Velocity data obtained from Figures 2 and 3)

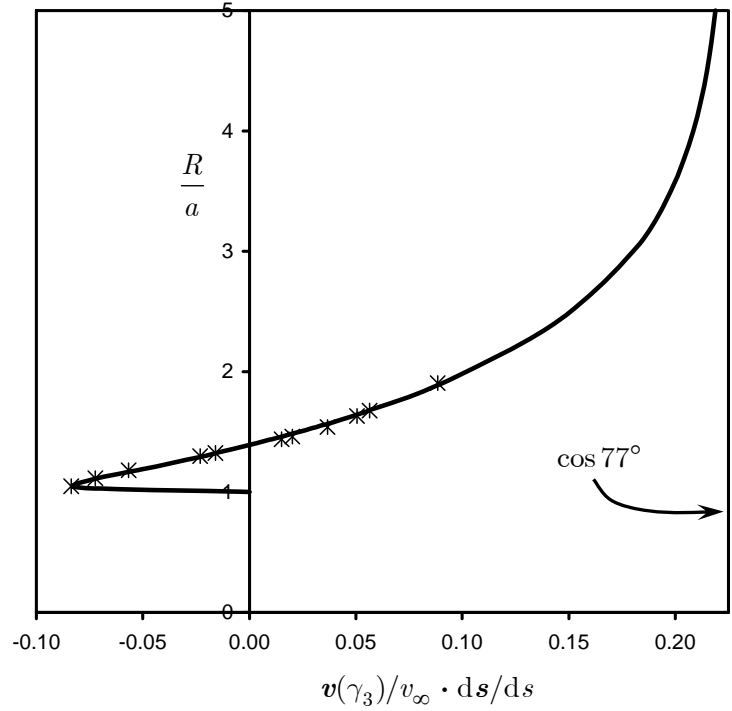

Figure 5. Components of the dimensionless mean velocity projected in the direction of the contour $\gamma_3$ with $\theta = 77°$. The points shown were taken from the projected velocities shown in Figure 4. Constants in (3): $c = 0.3326$, $m = 0.1003$, $k = 52.32$.

Integrating the terms in (4) gives:
$$\frac{\Gamma_{(1/2)}}{v_\infty} = \frac{-na}{n-1} - a\left(1 - \frac{1}{k}\right)\cos\theta - ac\left(\frac{1}{m} + \frac{1}{k}\right).$$

The circulation is negative, indicating that it is in the opposite sense of the integration contour. Substituting the values $\Gamma = 2\Gamma_{(1/2)}$ and $a = D/2$ gives:
$$\frac{\Gamma}{Dv_\infty} = \frac{-n}{n-1} - \left(1 - \frac{1}{k}\right)\cos\theta - c\left(\frac{1}{m} + \frac{1}{k}\right). \tag{5}$$

With $n = 2$ (ideal flow past a circular cylinder) and the constants fit to the data of Perrin *et al.* 2006 (given in the caption of Figure 5), the circulation predicted for a separation angle of $\theta = 77°$ is:
$$\frac{\Gamma}{Dv_\infty} = -2.56. \tag{6}$$

This agrees quite well with the experimental range of 2.07~2.55 reported by Cantwell and Coles (1983) as well as the values reported in Figure 19 of Tanaka and Murata (1986).

Application of this method to a normal flat plate would be particularly simple because in that case the separation point is known. The velocity upstream of the forward stagnation point in that case can be taken from Kirchhoff's solution to the normal flat plate (Kirchhoff 1869).

## 3. Discussion

At the point of separation on the surface, the tangential velocity should obey the relation (Schlichting 1979, p. 132):

$$\left(\frac{\partial v}{\partial r}\right)_{r=0} = 0. \tag{7}$$

This effect is not modeled by the velocity profiles used in this study; however, its contribution to the integral of (3) is very small owing to the smallness of the region that it affects.

Typical values of the mean separation point reported for a circular cylinder are $\theta = 82°$ for a laminar boundary layer and $\theta = 108°$ for a turbulent boundary layer (Schlichting 1979). However, as seen in the case above, for which Cantwell and Coles (1983) reported a value of $\theta = 77°$, these are not fixed values but depend on Reynolds number. The plot made by Cantwell and Coles of the mean pressure coefficient shows very large scatter. This is largely due to vortex shedding; the attached boundary layer actually *only passes through* $\theta = 77°$ and makes excursions far past and far below that value. During these excursions, the peak circulation is generated when the attached boundary layer on one side of the cylinder is longest, and hence the instantaneous separation angle largest.

Wu *et al.* (2004) quantified these periodic excursions for $(10 \leq Re \leq 275)$ in terms of high and low values of the separation angle at each Reynolds number (which will be referred to herein as $\theta_{s,Hi}$ and $\theta_{s,Lo}$, respectively), and found that their difference gradually increases to around $15°$ at $Re = 275$, where $\theta_{s,Lo} \approx 100°$ and $\theta_{s,Hi} \approx 115°$.

One observation might be in order regarding the pressure coefficient, defined as:

$$c_p \equiv \frac{(p - p_\infty)}{\frac{1}{2}\rho v_\infty}. \tag{8}$$

Substituting the well-known result from Bernoulli's equation

$$p = p_T - \tfrac{1}{2}\rho v_\infty$$

into (8) yields the familiar form:

$$c_p = 1 - \left(\frac{v}{v_\infty}\right)^2.$$

Evaluating the velocity on the surface by means of the solution for ideal flow around a cylinder, namely $v = -2v_\infty \sin\theta$, gives:

$$c_p = 1 - 4\sin^2\theta. \tag{9}$$

Therefore in ideal flow, the minimum in the pressure coefficient is $c_p = -3$, which agrees fairly well with the extremum in the scatter of Figure 9 of Cantwell and Coles (1983). The fact that the turbulent transition waves (Bloor 1964) had not reached the boundary layer in that study suggests that this scatter was due to alternating vortex shedding and, more generally, that the "loss" in pressure coefficient does not occur gradually and uniformly over the entire length of the boundary layer. Instead, the solution for ideal flow may remain fairly accurate during the periodic excursions of the attached boundary layer all the way to $108°$ or thereabouts. It appears to be the mathematical averaging process that makes the "loss" appear gradual in plots of surface pressure coefficient.

The angle of "separation" that is said to "occur" at around $\theta_s = 80°$ once vortex shedding is well underway ($Re > \mathrm{O}(10^3)$ for a cylinder), has no significance in the actual flow because it is only located by the inflection point of the *mean* pressure coefficient, which results from a mathematical averaging of pressure coefficients over a shedding cycle. Nothing physically happens at $\theta_s = 80°$ that does not also happen at $\theta = 79°$ and $\theta = 81°$, especially not flow separation. The actual separation angle, $\theta_{s,Hi}$, is probably always greater than 90° *for all* Reynolds numbers. What can be said of the "laminar" separation angle when vortex shedding is present is that the angle is unstable and alternates sides. This brings down the average to less than 90° when $Re > \mathrm{O}(10^3)$. Between each alternation, however, drag is high, and this raises the computed *mean* drag coefficient. However, the alternations can be eliminated by spinning the cylinder appropriately about its axis. Even for Reynolds numbers that are orders of magnitude lower than those of the so called 'drag crisis,' the drag coefficient of the cylinder can be brought down uniformly to $c_D \approx 0.5$ by rotating it (Swanson 1961). Figure 3 of Swanson (1961) suggests that, at least in the entire Reynolds number range reported, namely, $3.6 \times 10^4 \leq Re \leq 5 \times 10^5$, rotation eliminates the drag coefficient's otherwise drastic Reynolds number dependence. Most importantly, it does this by reducing the drag coefficient for every Reynolds number tested by Swanson nearly to the level of the drag minimum. In commenting on Prandtl's earlier work, Schlichting (1979, p. 43) argued that this is because "on the side where the wall and stream move in the same direction, separation is completely prevented. Moreover, on the side where the wall and stream move in opposite directions, separation is slight, so that on the whole it is possible to obtain a good experimental approximation to perfect flow with circulation and a large lift."

Another way of viewing this is that the rotation simply eliminates the alternation of the separation angle from side to side. Schlichting's statement of Prandtl's observation that with rotation, "separation is slight" could be restated as: "separation occurs at $\theta > 90°$." What may not be readily apparent is that even when the cylinder is not rotating, separation occurs at $\theta > 90°$, albeit on alternating sides of the cylinder.

Therefore, for a non-rotating cylinder, the separation point does not "move" to the rear of the cylinder during the drag crisis. Rather, the separation angle simply stops alternating from side to side (or else the level or duration of drag during the brief interval between alternations is drastically reduced), and this moves the *mathematical average* to the rear of the cylinder. It is mathematically impossible for the mean of a periodic process to coincide with its extremum. Therefore, the *mean* separation point cannot be where flow separation actually occurs, if shedding is periodic. In fact, it may be that never at any Reynolds number is the actual separation angle $\theta_{s,Hi}$ less than 90°.

## 4. Conclusion

Estimates of the total circulation generated by a circular cylinder are derived by integration of the velocity on a loop containing the far field and the cylinder surface along the attached boundary layer. The approaching free-stream velocity was assumed to stagnate according to ideal flow about a circular cylinder. The circulation predicted by this method is considerably smaller than that predicted by (1) and agrees quite well with experimental measurements. This may help explain the "loss" in circulation sometimes referred to when using (1). In particular, the closer agreement with observed values suggests that this anomalous loss does not occur. Flow separation probably never occurs on the forward side of a circular cylinder, but because flow separation alternates from side to side, the calculated *mean* separation angle is reduced below 90° when $Re > \mathrm{O}(10^3)$. Rotation of the cylinder at the proper rate eliminates the alternations and therefore reduces the *mean* drag.